\documentclass{aa}

\usepackage{graphicx,times}

\begin{document}

\msnr{ms1403}

\title{Where have all the black holes gone?}

\author{T.\ Beckert\inst{1}, W. J.\ Duschl\inst{2,1}}

\offprints{T.\ Beckert, Max-Planck-Institut f\"ur Radioastronomie,
Auf dem H\"ugel 69, 53121 Bonn, Germany,
tbeckert@mpifr-bonn.mge.de}

\mail{W.J.\ Duschl, Institut f\"ur Theoretische Astrophysik,
Tiergartenstr.\ 15, 69121 Heidelberg, Germany;
wjd@ita.uni-heidelberg.de}

\institute{Max-Planck-Institut f\"ur Radioastronomie, Auf dem
H\"ugel 69, 53121 Bonn, Germany \and Institut f\"ur Theoretische
Astrophysik, Tiergartenstra{\ss}e 15, 69121 Heidelberg, Germany}

\date{Received \dots ; accepted \dots}

\titlerunning{Where have all the black holes gone?}

\authorrunning{T.\ Beckert \& W.J. Duschl}

\date{Received / Accepted}

\abstract{We have calculated stationary models for accretion disks
around super-massive black holes in galactic nuclei. Our models
show that below a critical mass flow rate of $\sim
3\,10^{-3}\,\dot{M}_\mathrm{Edd}$ advection will dominate the
energy budget while above that rate all the viscously liberated
energy is radiated. The radiation efficiency declines steeply
below that critical rate. This leads to a clear dichotomy between
AGN and normal galaxies which is not so much given by differences
in the mass flow rate but by the radiation efficiency. At very low
mass accretion rates below $5\,10^{-5} \dot{M}_\mathrm{Edd}$
synchrotron emission and Bremsstrahlung dominate the SED, while
above $2\,10^{-4} \dot{M}_\mathrm{Edd}$ the inverse Compton
radiation from synchrotron seed photons produce flat to inverted
SEDs from the radio to X-rays. Finally we discuss the implications
of these findings for AGN duty cycles and the long-term AGN
evolution.
}

\maketitle

\section{Evidence for Black Holes and Advection-Dominated Accretion}
The existence of black holes (BH) in the centers of galaxies is
now widely accepted and the best mass determinations are known for
Sgr A$^*$ in the Galactic Center  ($M_\mathrm{BH} = 2.6\,10^6
M_\odot$) from the stellar velocity dispersion, and for NGC4258
from keplerian rotation of maser spots in an accretion disk.
Beside these very low luminosity AGNs, the masses of BH in quasars
have been estimated from their continuum luminosity and the
H$\beta$ line width (Laor \cite{L00}). It turns out, that
radio-loud quasars and radio galaxies, which show powerful jets,
are found in large elliptical galaxies with the most massive BHs
$\sim 10^9 M_\odot$.

 Seyfert Galaxies, on the other hand, are spiral galaxies
which host an AGN . Activity in the nucleus, which is powered by
accretion into a BH, can be discriminated against starbursts in
Seyferts from radio and X-ray observations. Radio cores and jets
with brightness temperatures above $10^8$K have been detected in
some Seyferts (Ulvestad et al. \cite{U99}; Mundell et al.
\cite{M00}; Falcke et al. \cite{F00}). Their flux stability over
several years exculdes radio supernovae as the power supply. The
X-ray emission shows rapid variability and in some cases a
redshifted Fe K$\alpha$ line, which is an indicator of
relativistic motion in the accretion disk around the BH. The
masses of BHs in some Seyfert 1 galaxies have been measured by
reverberation mapping of variable and correlated continuum and
line emission (Peterson \& Wandel \cite{PW00}). These measurements
are in reasonable agreement with the $M_\mathrm{BH}$--$\sigma$
relation of enclosed mass ($\sim M_\mathrm{BH}$) versus velocity
dispersions $\sigma$ in bulges of normal galaxies (Gebhardt et al.
\cite{Geb00}). It is therefore reasonable to assume the
existence of supermassive BH in most elliptical galaxies and
spirals with bulges.

While in the high luminosity objects both jet and accretion disk
can be identified in the spectrum, the situation is different in
less luminous AGNs like weak Seyfert Galaxies and LINERs. But even
here small scale jets are commonly found (Falcke et al.
\cite{F00}) and argue for the existence of BHs.  For instance
NGC4258 is an interesting transition object showing both an outer
irradiated thin accretion disk and a small scale radio jet. A
geometrically thin standard accretion disk close to the black hole
can not be identified, but the ionizing X-rays maybe produced at
the base of the jet, which can be identical with the proposed
advection-dominated accretion flow (ADAF) within $100
R_\mathrm{S}$ (Gammie, Narayan \& Blandford \cite{GNB99};
$R_\mathrm{S}$: Schwarzschild radius). At even lower luminosities
the Galactic Center (Sgr A$^*$) with a BH mass of $2.6 \,10^6
M_\odot$ (Genzel et al. \cite{G97}) is the only visible AGN with a
power output of $\approx 2\,10^{-10} L_\mathrm{Edd}$. A comparable
object in any other galaxy (spiral or elliptical) would not be
seen as an AGN. Assuming an spherical and adiabatic Bondi inflow,
cooled only by Bremsstrahlung, gives an radiation efficiency $\eta
= 9\,10^{-3}(L/L_\mathrm{Edd})$ (Frank, King \& Raine
\cite{FKR92}). Bremsstrahlung will be emitted in X-rays and the
{\it Chandra} detection (Baganoff et al. \cite{BBB01}) of
$L_\mathrm{X} \sim 2\,10^{33}$ erg s$^{-1}$ is consistent with a
mass accretion rate of $\dot{M} = 1.6\,10^{-6} M_\odot$ yr$^{-1}$
and an efficiency of $\eta = 2.2\,10^{-8}$.  The sub-mm
luminosity of Sgr A$^*$ is about 30 times larger than the X-ray
flux and makes Sgr A$^*$ a unique object. We will discuss a
specific ADAF model for Sgr A$^*$ in Sec. \ref{SGR_Sec}. The
Bondi flow faces at least two problems: it does not allow for any
possible angular momentum of the inflow and does not include
magnetic fields, which lead to synchrotron emission at radio
frequencies and synchrotron self-compton cooling. Both can be
accounted for in ADAF models. They provide a reasonable
explanation for the spectral energy distribution (SED) of Sgr
A$^*$ with a mass accretion of $\approx 10^{-6} M_\odot$ yr$^{-1}$
and a radiative efficiency of $10^{-5}$. Beside the basically
unresolved radio core of Sgr A$^*$, it is not possible to identify
a jet in the Galactic Center.

In this paper we will explore the hypothesis, that most of the
normal galaxies  without substantial  AGN activity contain
supermassive black holes $M > 10^6 M_\odot$ some of which have
been active during the quasar phase  $ 0.3 < z < 5$ ($z$ being the
cosmological redshift) and are quietly accreting in an ADAF mode
today. Spectral properties of ADAFs with rather large mass
accretion rates are explored in Sec. 2. The total luminosities and
spectral energy distributions are of interest for weak AGNs (Ho
\cite{H99}). We investigate the transition from standard thin disk
accretion to ADAFs and vice versa as an upper limit in the mass
accretion rate for ADAFs in Sec. \ref{limits}. The combined
consequences for accretion in normal galaxies are discussed in
Sec. \ref{disc_cons}.

\section{Radiation characteristics of ADAFs}
\begin{figure}
  \resizebox{\hsize}{!}{\includegraphics{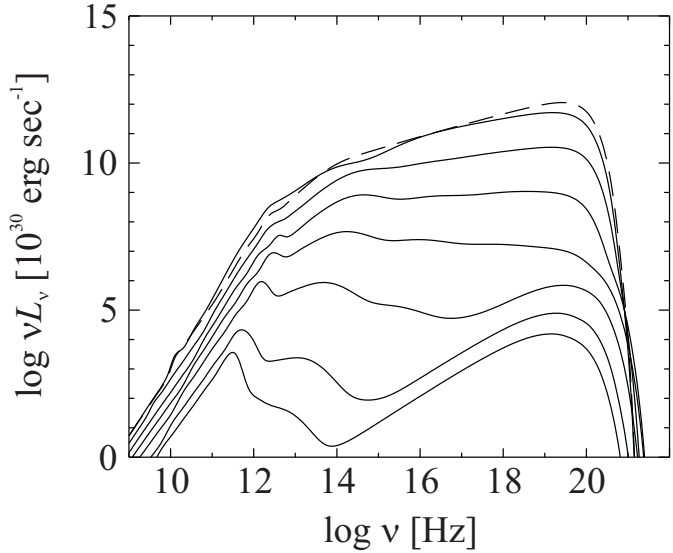}}
  \caption{Spectral energy distribution (SED) for ADAF models with a
  black hole mass of $2.6 \,10^6 M_\odot$. The mass accretion rates are
  $\dot{M} = (10^{-6},2\,10^{-6},5\,10^{-6},10^{-5},2\,10^{-5},5\,10^{-5},
  10^{-4},1.27\,10^{-4})\,M_\odot$ yr$^{-1}$.
  The scale-free accretion rate for this BH is $\dot{m} = 17.33 [\dot{M}/
  (M_\odot \mathrm{yr}^{-1})]$. The spectral energy flux increases for
  increasing mass accretion rate with an exception at $\dot{M}
  =1.27\,10^{-4}$, which is plotted with a dashed curve.\label{nulnu}}
\end{figure}
\begin{figure}
  \resizebox{\hsize}{!}{\includegraphics{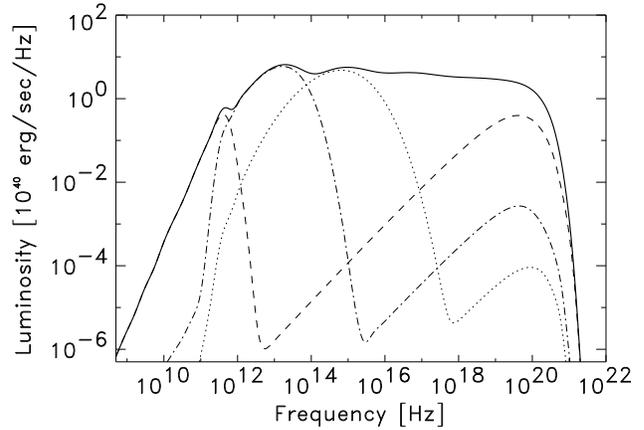}}
  \caption{\label{fig1a}
  SED for an ADAF around a $10^9 M_\odot$ black hole. The
  mass accretion rate is $\dot{m} = 3.6\,10^{-4}$ corresponding to
  $8\,10^{-3}  M_\odot$/yr. The seed photon flux is represented as dashed line,
  the first and second Compton humps are dash-dotted and dotted respectively.
  Further Comptonization is important but individual humps are smeared out.
  Only the total SED is shown as solid line. The dependence of the
  spectral shape on the black hole mass is weak, while the total flux scales
  with the mass accretion rate according to Eq. (\ref{effic}).}
\end{figure}
The calculation of emission spectra from ADAFs are based on
self-similar solutions for density, inflow velocity, rotation, and
ion temperature presented in Beckert (\cite{Beckert00}), which are
a generalisation of the Narayan \& Yi (\cite{NY94}) solution in
the Newtonian limit. Gravitational redshift and the small volume
of the general-relativistic region makes it reasonable to extend
the solution down to the black hole horizon at $R_\mathrm{S}$.
Aspects of the Kerr metrics are not considered. The electron
temperature is calculated from thermal balance between Coulomb
heating by ions, adiabatic compression, advective energy
transport, and radiation cooling. The primary radiation mechanisms
are Bremsstrahlung and synchrotron emission. Both are local in the
sense, that they only depend on the local state of the flow. At
large mass accretion rates multiple inverse Compton scattering
(IC) of synchrotron photons becomes the dominant cooling
mechanism. The photon flux to be scattered is produced at
different radii in the flow and is highly anisotropic, which makes
it a non-local process. Inverse Compton scattering and the
resulting cooling rate modify the density and temperature of the
ADAF and are treated iteratively. For calculating the IC
radiation, we used a method proposed by Poutanen \& Svensson
(\cite{PS96}) for thermal electrons, which we have modified to
allow for the quasi-spherical geometry of ADAFs. From the total
cooling we take the global radiation efficiency
$\varepsilon_\mathrm{Rad} = Q^-/Q^+$ to recalculate the
self-similar ADAF structure and electron temperature. We
then seek convergence of the assumed and posteriori
calculated $\varepsilon_{\mathrm{Rad}}$ to get globally consistent
ADAF models with correct radiation spectra. For the ADAF we assume
energy equipartition between ions and magnetic field. This
determines the synchrotron emission and provides additional
pressure to support the flow and lower the adiabatic index below
$5/3$ to make ADAFs possible at all. For the viscosity we use the
standard $\alpha$-prescription with $\alpha = 0.1$ well above the
critical value of $\alpha_{\mathrm{crit}} \approx 10^{-2}$ for the
transition to convection-dominated accretion flows (CDAFs)
described by Narayan, Igumenshchev \& Abramowicz (\cite{NIA00}).
No outflows, jets or wind infall as investigated in Beckert
(\cite{Beckert00}) are assumed, even though a wind infall is
suggested for the Galactic Center (Melia \& Coker \cite{MC99}).
The radial viscous break due to bulk and shear viscosity is
included, but it does not dominate the dynamical state of the flow
for $\alpha \le 0.1$.

From the described model we can construct spectral energy
distributions (SED) for different mass accretion rates. The mass
of the central BH has only a weak influence on the SED, which
is not included in the scaling of $\dot{M}$ to the Eddington
accretion rate, and is not considered here. We assume a mass of
$2.6\,10^6 M_\odot$ for the SED in Fig.\,\ref{nulnu}, appropriate for
the Galactic Center,  with a
corresponding Eddington limit of $\dot{M}_\mathrm{Edd} = 0.0577
M_\odot$ yr$^{-1}$ (in this paper we define the Eddington
accretion rate with an efficiency $\eta =
L_\mathrm{Edd}/(\dot{M}_\mathrm{Edd} c^2) = 0.1$) and use the
scale free accretion rate $\dot{m} = \dot{M}/\dot{M}_\mathrm{Edd}$
in the following discussion. The spectral luminosity in Fig.
\ref{nulnu} scales as the black hole mass, $\nu L_\nu \propto
M_\mathrm{BH}$. For comparision we show in Fig.\,\ref{fig1a} the SED
for an ADAF arround a $10^9 M_\odot$ black hole with $\dot{m} = 3.6\,10^{-4}$.
Fig.\,\ref{figeffic} demonsttrates that the radiation efficiency for
this flow is independend of the black hole mass.

The SEDs for $\dot{m}$ between $1.7\,10^{-5}$ and $2.2\,10^{-3}$
are shown in Fig.\,\ref{nulnu}. The presented model spectra are
accurate above 30 GHz, and they show that the synchrotron
emission rises in flux from $10^{33}$ to $10^{37}$ erg s$^{-1}$
and shifts in frequency from $2\,10^{11}$ Hz to $2\,10^{12}$ Hz at
$\dot{m} = 3.5\,10^{-4}$ and back to smaller frequency for larger
$\dot{m}$. Above $\dot{m} = 1.5\,10^{-4}$ the Thomson optical
depth for synchrotron photons from central regions around $3
R_\mathrm{S}$ is significant, and Compton scattering broadens the
synchrotron peak and make it less prominent, compared to the IC
emission. The dominating peak in the IC part of the spectrum,
which can be identified between the synchrotron and the
Bremsstrahlung peaks at $10^{19 \ldots 20}$ Hz, is the second
Compton peak of twice scattered synchrotron photons. The
synchrotron seed photons are produced in a region closer to
the BH than the Compton scattered radiation. So the seed photon
flux for first Compton scattering is anisotropic, and most
synchrotron photons are scattered back into a high density and
high temperature region with the largest optical depth. The second
Compton peak is therefore the dominant one, and the asymmetry
between even and odd scattering order decreases thereafter,
because the photon field to be scattered becomes more and more
isotropic. The SED becomes flat or inverted due to multiple IC
scattering above $10^{14}$ Hz for $\dot{m} \ge 3\,10^{-4}$. The
Bremsstrahlung peak will only be recognized below $\dot{m} <
1.5\,10^{-4}$. The peak does not shift very much in frequency as
the maximum of electron temperature only varies between
$2.5\,10^9$K and $8\,10^9$K, where the highest $T_{\mathrm{e}}$
are achieved at $\dot{m}  \approx 2\,10^{-4}$ very close to the
horizon and the photons from that radius are significantly
redshifted.

\begin{figure}
\resizebox{\hsize}{!}{\includegraphics{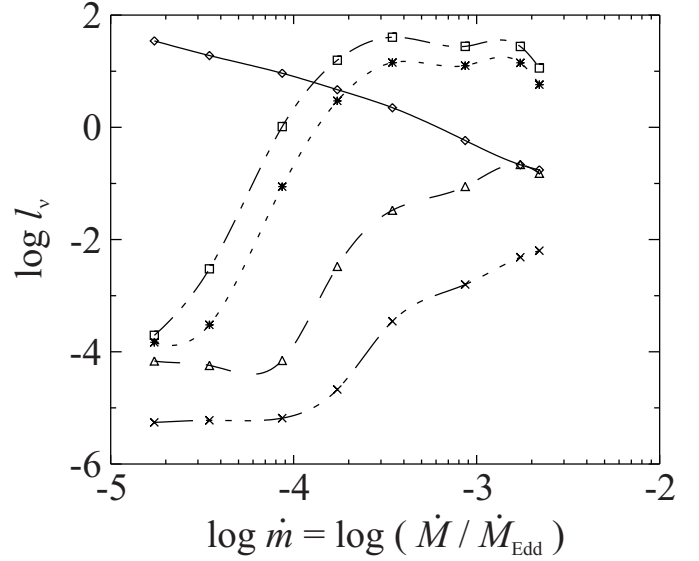}}
\caption{Normalized flux $l_\nu$ (see Eq. \ref{fluxL}) at 86 GHz
         (solid line), in K (dash-dotted)
         and V band (dotted), at 1 keV (dashed) and 100 keV
         (3$\times$dotted-dashed) as a function of mass accretion rate.
         The models are calculated for a $2.6\,10^6 M_\odot$ black hole.
         The efficiency $\eta$ in Eq. (\ref{fluxL}) is 10\% and the SED
         is assumed to peak at 100 keV. \label{fig2} }
\end{figure}

One major prediction of ADAF models is the different evolution of
observable flux in different frequency bands. This is expressed in
the scale-free spectral luminosity $l_\nu$. We scale the radiation
flux $L_\nu$ to the total, frequency integrated luminosity of the
specific model from Eq. (\ref{effic}) and define
\begin{equation} \label{fluxL}
   l_\nu =
   \frac{L_\nu}{\eta \dot{M}_\mathrm{Edd} c^2} \left[\frac{\dot{m}}
   {10^{-3}}\right]^{-2.3} \left[\frac{ h\nu_\mathrm{max}}
   { 100\,\mathrm{keV}}\right] \quad .
\end{equation}
We see in Fig.\ref{fig2} that the importance of synchrotron emission at 86
GHz decreases with rising $\dot{m}$, the X-rays follow the total
luminosity at first, and become more important when they are
dominated by IC emission above $\dot{m} \ge 2\,10^{-4}$. The
importance of IC emission is most prominently seen in K and V
band. The flux is rapidly rising with $\dot{m}$ and saturates
above $\dot{m} = 2\,10^{-4}$ when IC is dominating the cooling and
therefore IC follows the total luminosity. The described spectra
show only a weak dependence on the absolute mass of the BH between
$10^4$ and $10^9 M_\odot$, for which we have tested the models.
Larger BH masses imply smaller densities in the accretion flow at
the same $\dot{m}$. Consequently the electron temperature also
becomes smaller due to weaker Coulomb coupling. As the only
significant consequence for the SED, the position of the
synchrotron peak is anticorrelated with $M_\mathrm{BH}$, and it
shifts to smaller frequencies for larger BH masses.

 \section{A Note on the Galactic Center Source Sgr A$^*$}
\label{SGR_Sec}

\begin{figure}
  \resizebox{1.03\hsize}{!}{\includegraphics{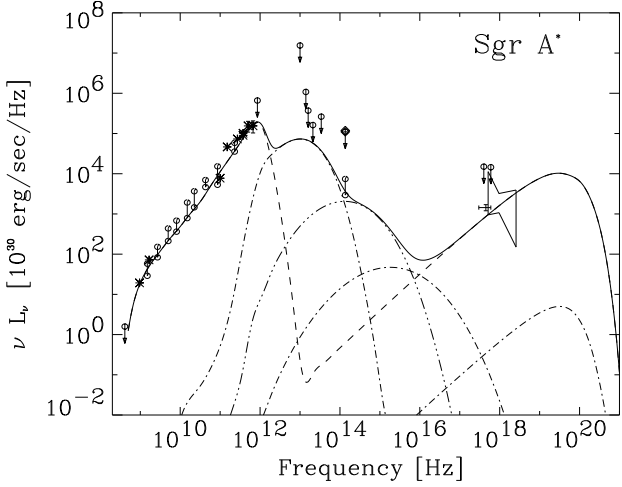}}
  \caption{ \label{SGRSED}
  Spectral energy distribution (SED) for an ADAF model with a
  wind infall appropriate for Sgr A$^*$. The black hole mass is
  $2.6 \,10^6 M_\odot$ and the mass accretion rate at the last stable
  orbit $\dot{M} = 9\,10^{-7} M_\odot$ yr$^{-1}$ corresponding to
  $\dot{m} = 1.6\,10^{-5}$. This special model is briefly described in
  Sec. \ref{SGR_Sec}. The dashed line shows the synchrotron and
  bremsstrahlung emission. The first three inverse Compton stages
  are given as dash-dotted lines.
  The total spectrum (thick line) has a remarkably flat radio
  slope $\nu L_\nu \propto \nu^{1.3}$ consistent with the observations.}
\end{figure}

The enigmatic radio source Sgr A$^*$ in the Galactic Center is
coincident with the center of gravity of an enclosed mass of
$2.6\,10^6 M_\odot$. It was considered to be one of the test cases
for ADAF models (Narayan et al. \cite{N98}), but the SED of the
source poses three problems for standard ADAFs. (1) The observed
radio spectrum is much flatter than predicted, so that only the
sub-mm bump (Falcke \cite{F99}) is nowadays attributed to the
accretion flow. Most of the radio emission at cm-wavelength must
then be produced by an outflow or jet (Falcke \& Markoff
\cite{FM00}). (2) The X-ray spectrum as derived from {\it Chandra}
observations (Baganoff et al. \cite{BMM01}) has a different slope
than expected from thermal bremsstrahlung coming from the ADAF.
(3) The observed rapid variability in X-rays (Baganoff et al.
\cite{BBB01}) restricts the size of the variable emitting region
to less than $10 R_\mathrm{S}$. The spectrum at high X-ray fluxes
is harder than at low flux levels. This can be explained by
inverse Compton emission of relativistic electrons in a jet
(Markoff et al. \cite{M01}) but even a jet has to be powered by an
accretion process and bremsstrahlung emission of the accreting gas
is unavoidable. In contrast to these recent scenarios, here we
present an ADAF-wind infall model, where the gas in the accretion
flow is heated by wind infall at all radii (Beckert,
\cite{Beckert00}) with a steeper density profile $\Sigma \propto
r^{-1/2-\beta}$ than normal ADAFs. The synchrotron emission
dominates the SED (Fig. \ref{SGRSED}) due to a strongly magnetised
ADAF $\beta_P = P_{\mathrm{Gas}}/P_{\mathrm{total}} < 0.5$.

The wind infall is assumed to be strong $\beta = 0.24$ and rotates
with $\Omega /2$ of the ADAF. The flow is magnetically dominated
with $\beta_P = 0.35$ and the radiative efficiency $\epsilon =
9\,10^{-5}$ is larger than in the other models presented in this
paper due to the larger electron temperature and the stronger
magnetic fields, which leads to increased synchrotron emission.
Viscosity is described by an $\alpha$-parametrisation with $\alpha
= 0.08$, lower than for the other ADAF models in Sec. 2, but
convection is still expected to be unimportant (Narayan,
Igumenshchev \& Abramowicz \cite{NIA00}). This model gives a good
fit to the radio spectrum but the problem with the X-ray
observation persists.

\section{Limits on the mass accretion rate\label{limits}}
\begin{figure}
\resizebox{\hsize}{!}{\includegraphics{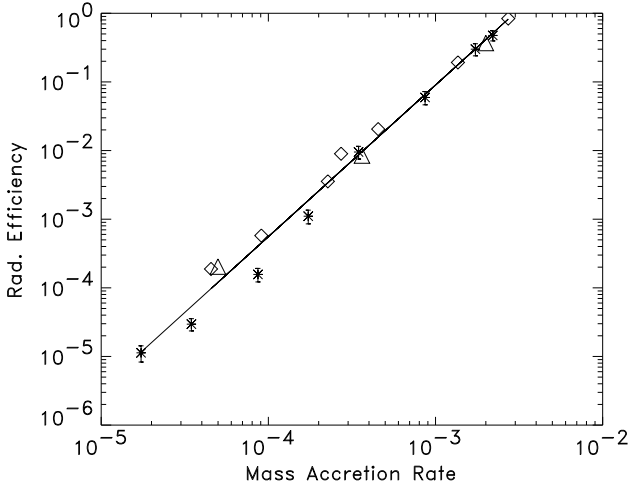}}
\caption{Radiation efficiency $\varepsilon$ as a function of mass accretion
         rate $\dot{m}$ with efficiency $\eta = 0.1$ to compare with the
         expectation for efficient radiation cooling.
         Model calculations are made for a  $2.62\,10^6 M_\odot$
          black hole (symbol: stars), for $10^8 M_\odot$ (diamonds), and for a
         $10^9 M_\odot$ (triangles) black hole.
         The estimated errors for the numerical convergence
         of assumed and posteriori derived $\varepsilon$ are given for
         the $2.6\,10^6 M_\odot$ black hole. The data are fitted
         with Eq. (\ref{effic}).\label{figeffic}}
\end{figure}

The luminosity of the synchrotron peak relative to Bremsstrahlung
and IC emission depends only on $\dot{m}$, and it is unaffected by
the absolute mass of the BH. The global radiative efficiency
$\varepsilon_\mathrm{Rad}$, defined in the previous section,
gives the cooling rate $Q^-$ with respect to the viscous heating
rate. The heating rate itself depends on
$\varepsilon_\mathrm{Rad}$ in the ADAF model, and it is therefore
not useful for comparing different models. In describing the
efficiency of an accretion flow, it is more reasonable to use
$\varepsilon = Q^-/(\eta \dot{M} c^2)$. In the following we assume
a standard efficiency of $\eta = 0.1$ . The quantity
$\varepsilon$ gives the ratio between the actual luminosity of the
accretion flow and the value expected for a thin and effectively
radiating accretion disk at the same mass accretion rate. The
radiative efficiency $\varepsilon$ is shown for our model
calculations in Fig. \ref{figeffic}. As described in the previous
section, different emission mechanisms dominate at different
accretion rates, and we do not expect $\varepsilon$ to follow a
simple power law in $\dot{m}$. The fit
\begin{equation} \label{effic}
  \varepsilon = \left(340\ \dot{m}\right)^{2.2}
\end{equation}
therefore does not represent a theoretically derived physical law,
and has to be taken with caution when used outside the scope
of the calculated models. Nonetheless the results allow us to
conclude that no consistent ADAF models with mass accretion rates
larger than $\dot{m}_\mathrm{crit} = 2.95\,10^{-3}$ are feasible.
The reason is that IC emission increases rapidly with electron
temperature and density. For increasing mass accretion rates both
density and surface density increase linearly as long as the
accretion velocity stays the same (this is true for ADAFs with
$\varepsilon \le 0.3$, which are consequently
advection-dominated). The maximum $T_\mathrm{e}$ in the flow
depends only weakly on $\dot{m}$ for IC dominated cooling. The
Thomson optical depth is
\begin{equation}
  \tau = \frac{3 \dot{m}}{\eta \alpha} \sqrt{\frac{R_\mathrm{S}}{2 r}}\quad ,
\end{equation}
and the mean energy which a photon gains from one scattering is $2
\gamma^2 h \nu \approx 6.5 h\nu$. The seed photons have $\nu
\approx 10^{12}$Hz, and upscattering stops if $(2\gamma^2)^n =
\gamma m_{\mathrm{e}}c^2/(h\nu)$, which results in photons gaining
energy in up to $n \approx 10$ consecutive scatterings. Starting
from a synchrotron luminosity $L_{\mathrm{s}}$ the IC luminosity
will be $ L_{\mathrm{IC}} \approx (2 \gamma^2 \tau)^n
L_\mathrm{s}$, if $2 \gamma^2 \tau > 1$, and the spectrum will be
flat for $2 \gamma^2 \tau = 1$. In the case of $\eta = \alpha =
0.1$  and $r \sim 3 R_\mathrm{S}$ this criterion is fulfilled, if
$\dot{m} = 1.3\,10^{-3}$. This is a reasonable order of magnitude
estimate, when compared to the model spectra described above. With
a constant synchrotron luminosity of $5\,10^{-7} L_\mathrm{Edd}$,
we find in the IC dominated regime
\begin{equation}
  L_{\mathrm{IC}} = 5\,10^{-7}\,(8\frac{\dot{m}}{\eta \alpha})^n\,
  L_\mathrm{Edd}\quad .
\end{equation}
An upper limit for the mass accretion rate is
\begin{equation}
 \dot{m}_\mathrm{c} = 0.53\,\eta\,\alpha ,
\end{equation}
which is in rough agreement with the upper limit found by fitting
the total luminosity in Eq. (\ref{effic}). It must be noted that
this limit $\dot{m}_\mathrm{c}$ has a different $\alpha$-dependence than
other limits $\dot{m}_\mathrm{crit} = \xi \alpha^2$
(Narayan \cite{N96}); Esin, McClintock, Narayan \cite{E97}) which give a wide
range for $\xi$ between 0.1 and 1.3 depending on details of the models.

Another local criterion for the existence of ADAFs is given by the
imbalance of Bremsstrahlung cooling and viscous heating. At large
radii the electron cooling rate, which is coupled to the ion
heating rate by the radiation efficiency,  decreases faster than
Coulomb coupling between thermal ions and electrons. The electron
temperature is therefore close to the ion temperature at
these radii. The ions are close to the viral temperature as long
as the radiative efficiency is significantly  smaller than 1,
which is the case for all calculated models here. In the region
where $T_\mathrm{e}$ and $T_\mathrm{ion}$ are equal, the
Bremsstrahlung cooling, which dominates at large radii, decreases
as $r^{-5/2}$, while the viscous heating falls off as $r^{-3}$. At
a critical outer radius, the radiative efficiency is 1 and no ADAF
is possible at larger radii. This outer radius is found to be
\begin{equation} \label{BremsR}
  R_\mathrm{out} = 5\,10^2 \alpha^4\,\dot{m}^{-1}\,R_\mathrm{S} \quad ,
\end{equation}
which is valid for $\alpha \le 0.1$ and  means that for $\dot{m} >
0.13$  Bremsstrahlung from an ADAF is more efficient than the
viscous heating outside the marginally stable orbit at $R =
3R_\mathrm{S}$. From studies of the Galactic Center we know
that no standard thin accretion disk exists within $10^5
R_\mathrm{S}$ and the mass accretion rate from spectral fitting of
an ADAF model gives $\dot{m} \approx 1.6\,10^{-5}$. From Eq.
(\ref{BremsR}) this results in a lower limit estimate of $\alpha >
0.01$.

\section{Discussion -- Consequences for AGN Evolution\label{disc_cons}}
We find that the radiation efficiency depends strongly on the mass
accretion rate ($\propto {\dot m}^{2.3}$) below a certain critical value
$\dot{m}_\mathrm{crit}$ of the mass accretion rate (in units of its
Eddington value). Above $\dot{m}_\mathrm{crit}$, $\varepsilon = 1$.
In other words, below the critical mass
accretion rate, the efficiency decreases so rapidly that
one can discern two regimes:
\begin{equation}
  \label{eq:approxeps}
  \varepsilon = \left\{ \begin{array}{ll} 1 & \mathrm{for\ }\dot m >
  {\dot m}_\mathrm{crit}\\ 0 & \mathrm{for\ }\dot m < {\dot
  m}_\mathrm{crit}\\ \end{array} \right.,
\end{equation}
with only a small transition zone in $\dot{m}$ below ${\dot m}_\mathrm{crit}$
where $\varepsilon$ differs significantly from 0. For the purpose of our
present discussion, however, the approximation of Eq.\
(\ref{eq:approxeps})\ suffices.

Furthermore, our numerical models show that the relation outlined by
Eq.\ (\ref{eq:approxeps})\ holds for the entire range of $M_\mathrm{BH}$
investigated. In the following we will use the value ${\dot m}_\mathrm{crit}
= 0.003$ as derived by interpolating our numerical models (Eq.\ \ref{effic}).

\subsection{The onset of nuclear activity}

The existence of ${\dot m}_\mathrm{crit}$ with the properties discussed
above translates into a fairly sharp transition between an
active and an inactive state of a galaxy. As soon as $\dot m$
falls below ${\dot m}_\mathrm{crit}$ the radiation efficiency of
the accretion decreases dramatically. In other words, already a
relatively small change in the mass flow rate around ${\dot
m}_\mathrm{crit}$ suffices to ``switch off" an AGN, and vice
versa. The difference between a {\it normal\/} and an {\it
active\/} galaxy is then not due to a difference in $\dot m$ which
is as large as the difference in luminosities between the two
classes. A much more important reason is the steep decline in the
radiation efficiency for the accretion rates below which the disk turns
advection-dominated.

Our numerical models predict ${\dot m}_\mathrm{crit} \sim 0.003$.
This is in good agreement with observations (e.g., Peterson \&
Wandel \cite{PW00}, who find AGN only in the luminosity range
between $\sim 10^{-3}$ and 1 of the Eddington luminosity,
or---equivalently---the Eddington mass accretion rate). In our
interpretation the lack of galactic nuclei below
$10^{-3}\,L_\mathrm{Edd}$ is not due to a lack of galaxies with
mass accretion rates below $10^{-3}\,{\dot M}_\mathrm{Edd}$ but
rather due to the steep decline of radiation efficiency below this
critical value.

\subsection{The general properties of the evolution of AGN and of
their black holes}

For a supermassive black hole of mass $M_\mathrm{BH}$, one can
give an average accretion rate $\overline{\dot M}$ over its age
$\tau_\mathrm{BH}$ of
\begin{equation}
\overline{\dot M} \le \frac{M_\mathrm{BH}}{\tau_\mathrm{BH}}
\end{equation}
We can only give an upper limit of $\overline{\dot M}$, because there
is the possibility of a non-negligible seed mass of the black hole
which is not due to this type of accretion processes.

If, in addition, we assume, that the age of the BH is not very
much shorter than the age of its host galaxy and thus the Hubble time
(at the location of the black hole), $\tau_\mathrm{H}$, we can
write
\begin{equation}
\overline{\dot M} \le \frac{M_\mathrm{BH}}{\tau_\mathrm{H}}
\end{equation}
In the following, we express the time in units of $10^{10}\,$yr, and
the mass flow rates in units of the Eddington rate. We introduce the
abbreviations ${\dot m} = \dot M / {\dot M}_\mathrm{Edd}$, and
$\tau_{10} = \tau / 10^{10}\,$yrs, and get an average mass
accretion rate of
\begin{equation}
\overline{\dot m} \le 0.04 \eta \frac{1}{\tau_{10}}.
\end{equation}
In terms of the Eddington accretion rate, the average accretion
rate $\overline{\dot m}$ is independent of the accreting black
hole's mass. For a constant accretion efficiency $\eta$, it is
$\overline{\dot m} \propto t^{-1}_\mathrm{BH}$
a function of the age of the black hole,
or---for that matter---the Universe. This leads to two interesting
consequences:
\begin{itemize}
\item In terms of the Eddington rate, the accretion rate declines
as the black hole, the galaxy, and the Universe as a whole evolve;
\item The present-day average accretion rate is around $10^{-3}$ of
its Eddington value.
\end{itemize}

The present-day average accretion rate, $\overline{\dot m}_{(0)}$
is fairly close to the critical accretion rate ${\dot
m}_\mathrm{crit} \sim 0.003$ given by $\varepsilon = 1$ (Eq.\
\ref{effic}). For $\dot m > {\dot m}_\mathrm{crit}$ the radiation
efficiency is unity, while for smaller $\dot m$ it decreases
sharply $\propto m^{2.3}$. This means that relatively small
changes in the momentary accretion rate are capable of
transferring a galaxy from an inactive to an active state and vice
versa. In the course of the further evolution of the Universe, it
will become harder and harder for galaxies to turn active as the
average accretion rate (in Eddington units) becomes smaller and
smaller. This is compounded with and strengthened by a
diminishing supply of material available for accretion, i.e., by
an additional decrease of the absolute value of the average
accretion rate. Extrapolating back to earlier cosmological
epochs, we find (because of smaller $\tau_\mathrm{BH}$ and  higher
$\overline{\dot m}$) that  the likelihood for a galaxy to be
active was higher on two grounds: {\it (1)\/} The larger
$\overline{\dot m}$ the smaller a fluctuation suffices to turn a
galaxy active. {\it (2)\/} The further we go back in the evolution
of the Universe, the larger was the supply of gas available for
accretion.

\subsection{Constraints on the AGN duty cycle}

As discussed above, the mass $M_\mathrm{BH}$ and age
$\tau_\mathrm{BH}$ of a black hole define an upper limit for its
average mass accretion rate. During phases of activity, the mass
flow rate must be larger than $\overline{\dot m}$. Let us---for
the purpose of a crude estimate---assume that we have only two
states, namely the {\it AGN\/} phase, characterized by a mass flow
rate ${\dot m}_\mathrm{AGN} > {\dot m}_\mathrm{crit}$ lasting for
a period of time of $\theta \tau_\mathrm{H}$, and a {\it normal
galaxy\/} phase for which the mass flow rate ${\dot
m}_\mathrm{normal} < {\dot m}_\mathrm{crit}$ is correspondingly
smaller so as to maintain the average value $\overline{\dot m}$.
Let us, moreover, assume\footnote{This assumption is not necessary
for activity to set in, as we have seen in the previous Sect. For
the following, however, it is a handy assumption which does not
influence the results of our order-of-magnitude estimates.}
that $(1-\theta) {\dot m}_\mathrm{normal} \ll \theta {\dot
m}_\mathrm{AGN}$, then we get for the duty cycle
\begin{equation}
  \theta = \frac{\overline{\dot m}-{\dot m}_\mathrm{normal}}
  {{\dot m}_\mathrm{AGN}-{\dot m}_\mathrm{normal}} \geq
  \frac{\overline{\dot m}}{2{\dot m}_\mathrm{AGN}}.
\end{equation}

${\dot m}_\mathrm{AGN}\ge {\dot m}_\mathrm{crit}$ needs to be
fulfilled for an activity phase to occur at all. The stronger the
activity of a galaxy is the shorter it can be maintained. For
instance, for an AGN operating at its Eddington limit, i.e., at
${\dot m}_\mathrm{AGN} = 1$ this means that it cannot stay at this
level of activity---integrated over all individual phases of
activity---for longer than a fraction $\theta$ of its entire
evolution, i.e., some $10^7\,$yr in the present-day Universe. A
super-Eddington activity level can be maintained only for an even
shorter period of time.

 Derivation of a more detailed luminosity evolution of an AGN sample
requires, however, a treatment more detailed than the above
order-of-magnitude estimates. In particular it has to be
investigated whether real-world galaxies can maintain a
sufficiently high supply of matter for the accretion process over long
enough a period of time. This then involves, for instance, questions about
the accretion time scales and the long-term development of mass
reservoirs. This topic, however, is beyond the scope of the
present paper and will be addressed separately (Duschl \&\
Strittmatter, in prep.)

\section{Conclusions}
We have shown that advection-dominated accretion flows into black
holes display changing spectral energy distributions for different
mass accretion rates. At very low mass accretion rates below
$\dot{M} = 5\,10^{-5} \dot{M}_\mathrm{Edd}$ synchrotron emission
and Bremsstrahlung dominate the SED. Above $\dot{M} = 2\,10^{-4}
\dot{M}_\mathrm{Edd}$ the inverse Compton radiation from
synchrotron seed photons produce flat to inverted SEDs from the
radio to X-ray bands. Multiple inverse Compton scattering is the
most relevant cooling process for mildly relativistic electrons in
two-temperature ADAFs, whenever the actual radiation efficiency in
eq. (\ref{effic}) is larger than 0.07 or the total luminosity
larger than $4\,10^{-7} L_\mathrm{Edd}$. The rapidly increasing
cooling efficiency of the inverse Compton process sets an upper
limit for the mass accretion rate of ${\dot m}_\mathrm{crit} =
3\,10^{-3}$ for ADAFs. At larger mass accretion rates radiative
cooling balances the local viscous heating and advective energy
transport is unimportant. The resulting flows are less hot,
disk-like, and therefore much denser than ADAFs. Below the
critical accretion rate, the radiation efficiency declines rapidly
$\varepsilon \propto \dot{m}^{2.3}$ and the total luminosity is
\begin{equation}
  L_\mathrm{ADAF} = 3.45\,10^{-3}\,(\dot{m}/0.003)^{3.3}\,L_\mathrm{Edd}
  \quad .
\end{equation}

In addition to the changing spectral behavior, the combination of
a transition value between ADAFs and normal accretion disks
(${\dot m}_\mathrm{crit} \sim 3\,10^{-3})$, and the very steep
decline of the ADAF radiation efficiency below this transition
value towards smaller accretion rates ($\varepsilon \propto {\dot
m}^{2.3}$) leads to an apparent  dichotomy in what one
observes: Above ${\dot m}_\mathrm{crit}$, one will recognize the
sources as AGN; while below ${\dot m}_\mathrm{crit}$, practically
no nuclear activity is observable. In other words, accretion into
a super-massive black hole should be prominently visible only at
mass flow rates above ${\dot m}_\mathrm{crit}$, which is in good
agreement with observed AGN distributions. The major difference
between AGN and normal galaxies is not so much a difference in
mass flow rate---at least not by as much as the difference in
luminosities may make us think---but rather one in the radiative
efficiency.

At the same time, this means that---at accretion rates around
${\dot m}_\mathrm{crit}$---small changes in the mass flow rate are
sufficient to cause a strong difference in radiation efficiency
and thus nuclear luminosity. In other words, the crossing of ${\dot
m}_\mathrm{crit}$ acts almost like a {\it switch\/} which turns
AGNs on and off.

Finally, the combination of the black holes' masses and the mass
accretion rate allowed us to put constraints on the duty cycle of
AGN. It turned out that the most active AGN can maintain this
level of activity only for rather short time scales (of the order
of some $10^7$\,years).

\begin{acknowledgements}
We wish to thank the referee, Dr. Suzy Collin, for her very
helpful report on this paper. This work was in part supported by
the {\it Deutsche Forschungsgemeinschaft, DFG,\/} through grant
SFB439/C2.
\end{acknowledgements}

\end{document}